\documentclass[11pt,twocolumn,english]{scrartcl}
\usepackage{mathpazo}

\usepackage[T1]{fontenc}
\usepackage[latin9]{inputenc}
\usepackage[letterpaper]{geometry}
\usepackage{textcomp}
\usepackage{graphicx}

\makeatletter
\newcommand{\lyxaddress}[1]{
\par {\raggedright #1
\vspace{1.4em}
\noindent\par}
}

\@ifundefined{date}{}{\date{}}
\usepackage{numprint}
\usepackage[super]{nth}
\usepackage[superscript,biblabel]{cite}
\usepackage[font={sf,small}, labelfont={bf}, indention=0cm, format=plain]{caption}

\let\originalleft\left
\let\originalright\right
\renewcommand{\left}{\mathopen{}\mathclose\bgroup\originalleft}
\renewcommand{\right}{\aftergroup\egroup\originalright}

\makeatother

\usepackage{babel}
\begin{document}

\title{\textit{In vivo} optical imaging of physiological responses to photostimulation
in human photoreceptors}

\author{Dierck Hillmann,\textsuperscript{1,{*}} Hendrik Spahr,\textsuperscript{2,3}
Clara Pfäffle,\textsuperscript{2,3} \\
Helge Sudkamp,\textsuperscript{2,3} Gesa Franke,\textsuperscript{2,3}
Gereon Hüttmann\textsuperscript{2,3,4}}

\maketitle

\lyxaddress{\textsuperscript{1}Thorlabs GmbH, \\
Maria-Goeppert-Straße~9, 23562~Lübeck, Germany}

\lyxaddress{\textsuperscript{2}Institute of Biomedical Optics Lübeck, \\
Peter-Monnik-Weg~4, 23562~Lübeck, Germany}

\lyxaddress{\textsuperscript{3}Medical Laser Center Lübeck GmbH, \\
Peter-Monnik-Weg~4, 23562~Lübeck, Germany}

\lyxaddress{\textsuperscript{4}Airway Research Center North (ARCN), \\
Member of the German Center for Lung Research (DZL), Germany}

\lyxaddress{\textsuperscript{{*}}dhillmann@thorlabs.com}

\textsf{\textbf{Non-invasive functional imaging of molecular and cellular
processes of vision is expected to have immense impact on research
and clinical diagnostics. Although suitable intrinsic optical signals
(IOS) have been observed }}\textsf{\textbf{\textit{ex vivo}}}\textsf{\textbf{
and in immobilized animals }}\textsf{\textbf{\textit{in vivo}}}\textsf{\textbf{,
it was so far not possible to obtain convincing IOS of photoreceptor
activity in humans }}\textsf{\textbf{\textit{in vivo}}}\textsf{\textbf{.
Here, we observed spatially and temporally clearly resolved changes
in the optical path length of the photoreceptor outer segment as response
to an optical stimulus in living human. To obtain these changes, we
evaluated phase data of a parallelized and computationally aberration-corrected
optical coherence tomography (OCT) system. The non-invasive detection
of optical path length changes shows the neuronal photoreceptor activity
of single cones in living human retina, and, more importantly, it
provides a new diagnostic option in ophthalmology and neurology and
could give new insights into visual phototransduction in humans.}}

Spatially resolved optical detection of retinal function in living
humans promises new diagnostic possibilities and new insights into
\textit{in vivo} phototransduction. Vision starts with the photoabsorption
by the ``visual purple'' rhodopsin or an equivalent photopsin in
retinal photoreceptors, which triggers an amplifying biochemical chain
reaction finally leading to hyperpolarization of the cell and signaling
to the brain -- a process known as visual phototransduction. In addition
to initial bleaching of rhodopsin itself, several optical changes
arise during this vision process (e.g., \cite{Liebman:74,Harary:1978})
and in the past decades, this led scientists to investigate intrinsic
optical physiological responses to retinal photostimulation \textit{ex
vivo }and in animals \textit{in vivo} by various optical techniques,
e.g., microscopes, fundus cameras, and laser scanning ophthalmoscopes
(reviewed in \cite{Yao:2015}). In fact, different fast intrinsic
optical signals after light stimulus were reported, visible as changes
in light scattering, absorption, or birefringence, with time constants
from milliseconds to seconds. The best \textit{in vivo} results have
been weak signals, obtained by merely looking at the backscattering
intensity. In humans, however, these signals were disturbed by rapid
eye motion \cite{Jonnal:07,Grieve:2008,Srinivasan:09,Schmoll:2010,Teussink:15,Yao:2015}
and could therefore not be reliably detected. 

Principally, optical coherence tomography (OCT\cite{Huang:1991})
offers unique advantages for the observation of intrinsic optical
signals, as it provides three-dimensional infrared imaging of backscattered
light, and thereby achieves precise signal localization and minimizes
additional undesired photostimulation. Almost ten years ago weak changes
in backscattering intensity after retinal stimulation were observed
using OCT \textit{in vitro} \cite{Bizheva:2006-3693} and in animals
\cite{Srinivasan:06}. As with other imaging techniques, published
studies in humans had significant problems with motion, resulting
in even weaker signals, and were not able to convincingly demonstrate
the measurement of intrinsic optical signals \cite{Srinivasan:09,Schmoll:2010,Teussink:15}. 

\begin{figure}
\begin{centering}
\includegraphics{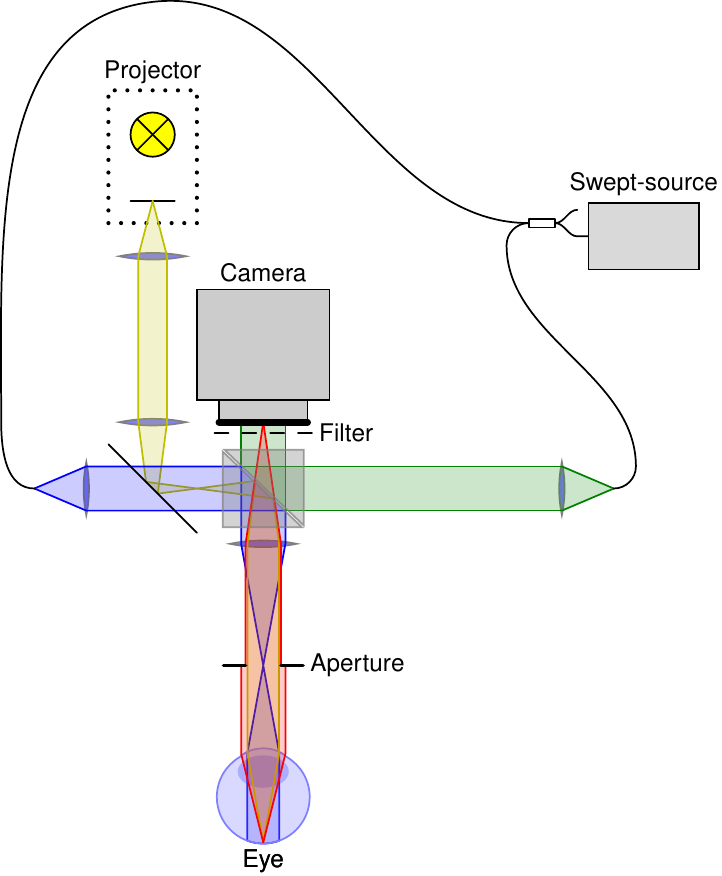}
\par\end{centering}

\caption{\textbf{\label{fig:Setup}Setup used to observe intrinsic optical
signals. }Light from a tunable laser source is split in two beams
for reference irradiation (green) and illumination of the retina (blue).
Backscattered light (red) is imaged onto an a fast CMOS camera, where
it is superimposed with the reference light. A projector with an LED
source is used to stimulate the retina (yellow). }
\end{figure}
In addition to backscattering intensity, state-of-the-art Fourier-domain
OCT (FD-OCT) also detects phases, which are sensitive to small changes
of the optical path length that the backscattered light has travelled.
The optical path length of the photoreceptor outer segment is proportional
to the time light requires to pass the cells, and it reflects morphology
and refractive index, which principally both could be affected during
phototransduction. Still, so far no study concentrated on the phase
signal to observe fast intrinsic optical signals: to achieve sufficient
motion stability for a detection of small phase changes \textit{in
vivo}, the requirements in imaging speed are enormous. Fully parallelized
OCT techniques, such as holoscopy \cite{Hillmann:11} and full-field
swept-source OCT (FF-SS-OCT) \cite{Povazay:06,Bonin:10}, can principally
acquire data significantly faster. Recently, FF-SS-OCT thereby detected
heart-beat induced pressure waves within retinal vessels \cite{Spahr:15}
and imaged single photoreceptors by using computational aberration
correction \cite{Adie:12,Shemonski:15,Hillmann:2015}.

\begin{figure}
\begin{centering}
\includegraphics{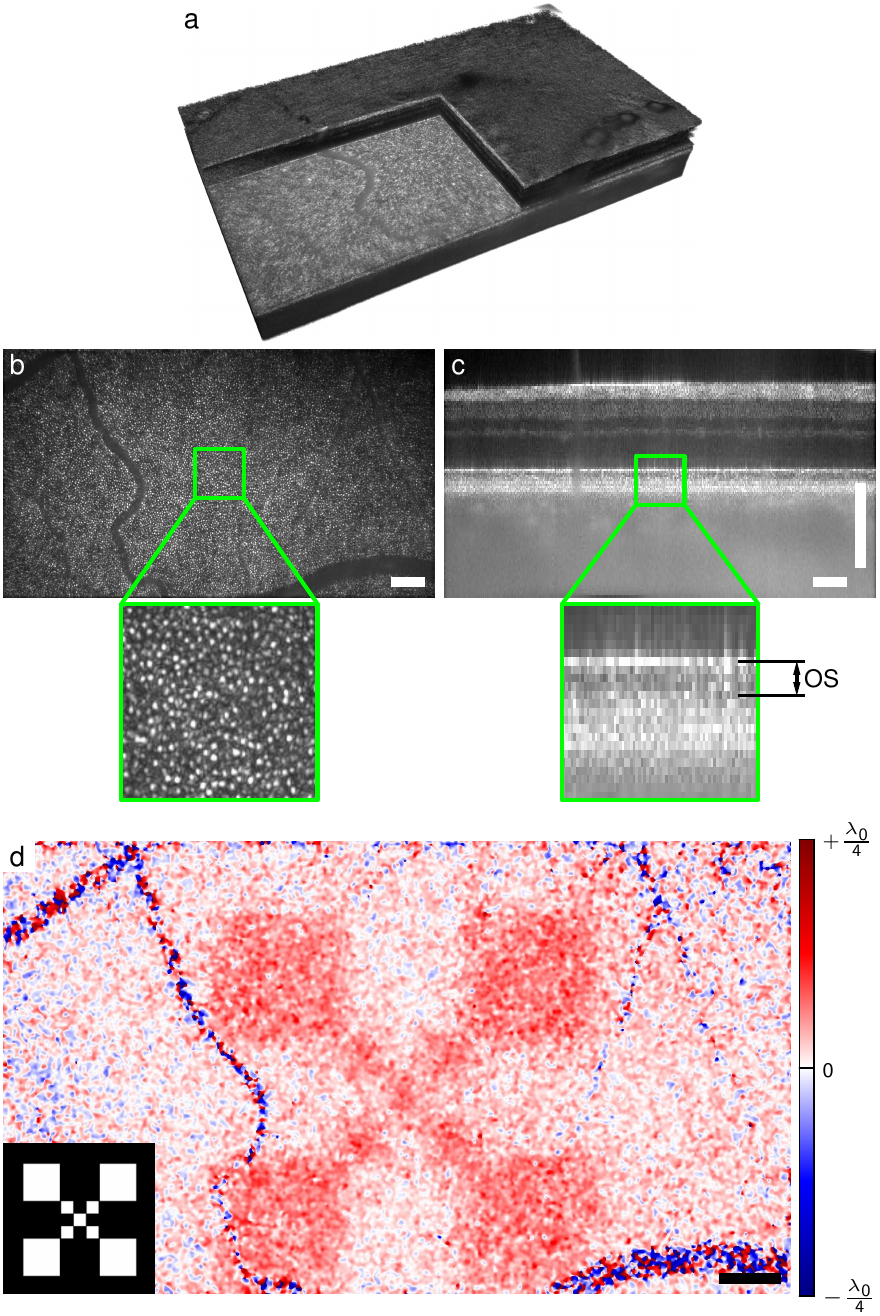}
\par\end{centering}

\caption{\textbf{\label{fig:TestStimulation}Retinal imaging and response to
optical stimulus.} a)~Volume of retina acquired by full-field swept-source
OCT. b)~En-face plane showing the backscattering intensity from the
outer segment. Cone photoreceptors are clearly visible. c)~Cross-sectional
view (B-scan) from the center of the recorded volume. Changes of the
optical path length within the outer segment (OS) were evaluated in
this study. d)~Spatially resolved change of the optical path length
$\Delta\ell$, 297~ms after the stimulus pattern, which is shown
in lower left inset. Response is high in the stimulated region and
random in blood vessels. Noise was reduced by a lateral Gaussian filter.
The center wavelength $\lambda_{0}$ of the OCT was 841~nm. Scale
bars are 200~\textmu m. }
\end{figure}
Here, we measured changes of the optical path length $\Delta\ell$
within the photoreceptor outer segment after projecting a 50~ms white
light pattern with a total radiant flux of 10~\textmu W onto the
retina of a healthy young adult. For this, we imaged the retina before,
during, and after the stimulus using a FF-SS-OCT system (Fig.~\ref{fig:Setup})
with up to 180~volumes/s -- significantly faster than other OCT systems
in research. We then reconstructed the acquired volumes, corrected
them for motion, segmented the inner segment/outer segment layer,
and flattened it. Figure~\ref{fig:TestStimulation}a shows a resulting
imaged volume from the periphery at about 11\textdegree . After correcting
the data numerically for ocular aberrations, cone photoreceptors in
the en face images (Fig~\ref{fig:TestStimulation}b) become clearly
visible. When evaluating phase difference changes between the photoreceptor
inner segment/outer segment junction and the outer segment tips (see
sectional view (B-scan) in Fig.~\ref{fig:TestStimulation}c), we
observed an increase of about 50~nm in the optical path length within
the outer segment, 297~ms after beginning of the stimulus (Fig.~\ref{fig:TestStimulation}d).
The signal is clearly related to the stimulus as spatial correspondence
to the projected pattern was obvious (bottom left inset of Fig.~\ref{fig:TestStimulation}d).
However, since the optical path length is the product of the refractive
index and the geometric length, we cannot differentiate between chemical
and structural changes. 

\begin{figure*}
\begin{centering}
\includegraphics{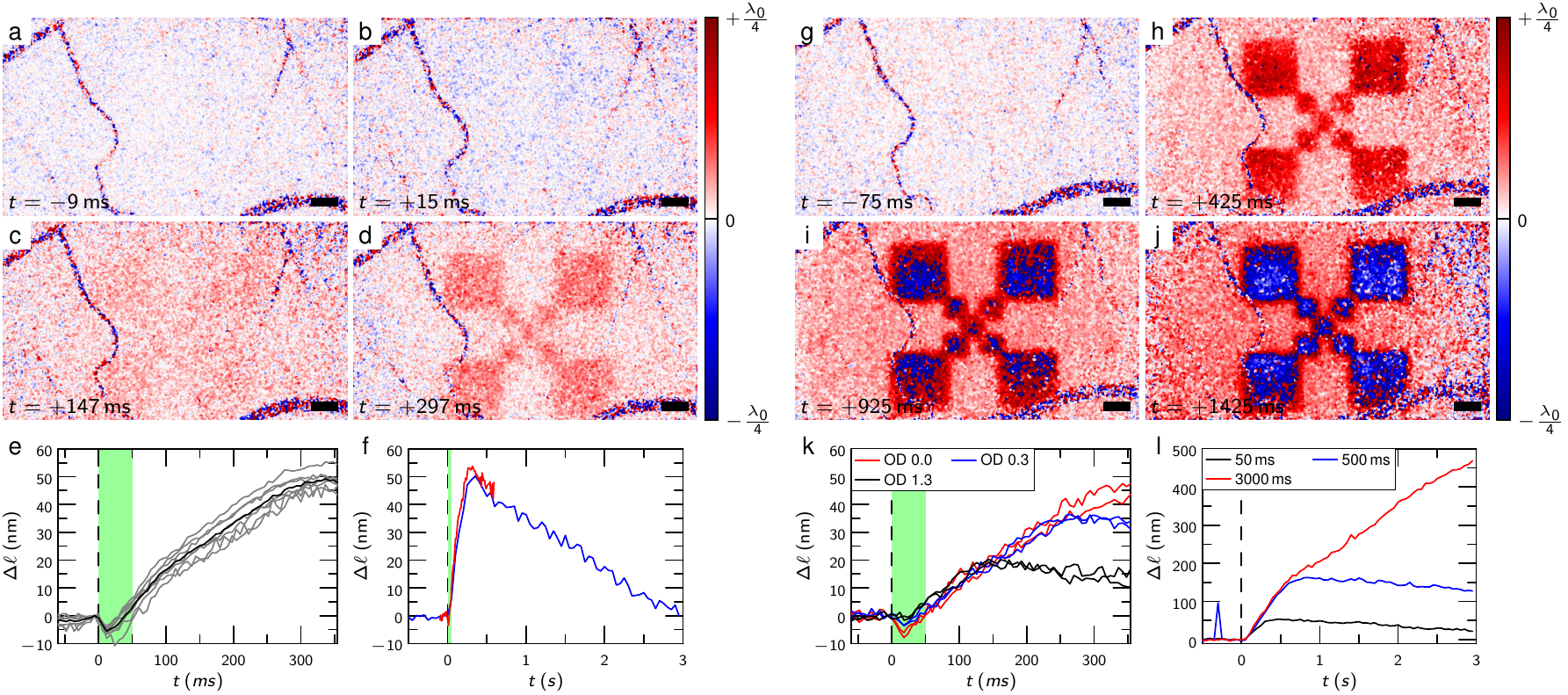}
\par\end{centering}

\caption{\textbf{\label{fig:Response}Timecourse of the intrinsic optical signal
$\Delta\ell$ for different stimuli. }a-d)~Response to a 50~ms stimulus;
a small negative response after about 17~ms (b) converts to a strong
positive stimulation (c-d). e,f)~Timecourse, averaged over the stimulated
area. The black line (e) is the average of 7 individual measurements
(gray lines). g-j)~Response to a permanent stimulus. An increasing
response is observed, finally leading to phase wrapping. k)~Response
to different flash intensities attenuated by OD 0.3 and OD 1.3, respectively.
l)~Response for 50~ms, 500~ms, and 3000~ms stimuli with 10~\textmu W.
The dashed black lines indicate the start of the stimulus at $t=0$
(e,f,k,l), the green areas (e,f,k) indicate the duration of the stimuli.
The center wavelength $\lambda_{0}$ of the OCT was 841~nm. Scale
bars are 200~\textmu m. }
\end{figure*}
We derived the timecourse of the signals from a series of 70 consecutive
volumes acquired at a rate of 180~Hz to provide a temporal resolution
of 6~ms and another series acquired at 20~volumes/s for a measurement
of more than 3~s, respectively; in both cases the 50~ms stimulus
started at the beginning of the \nth{10} volume. After averaging
the stimulated area (Fig.~\ref{fig:Response}e,f), we observed the
optical path length of the outer segments $\Delta\ell$ to decrease
by 5~nm during the first 15~ms of the stimulus. However, the signal
was weak and required averaging of 7 measurements to become clearly
visible. After these 15~ms, the optical path length started to increase
significantly, even several hundred milliseconds beyond the end of
the stimulus (Fig.~\ref{fig:Response}e). After about 300~ms the
signal reaches its maximum of 40~nm, then dropped for 2.5~s with
almost constant rate back to the ground level (Fig.~\ref{fig:Response}f).
Figures~\ref{fig:Response}a-d show images from this time series.

Bleaching of rhodopsin and activation of metarhodopsin II happens
in the order of a few milliseconds \cite{Liebman:74}, and the initial
response might therefore be related to the direct photoactivation.
In \textit{ex vivo }experiments with bovine rod outer segments, rapid
disc shrinkage after light exposure was in fact previously reported
\cite{Uhl1977}. The following increase of the optical path length
within the outer segment lasts significantly longer than the stimulus,
which indicates the signal follows activation of metarhodopsin II
and therefore might be related to its deactivation, follows the activation
of transducin, or is related to other secondary processes. Non-physiologic
effects, such as a thermal expansion, would cause a direct decrease
once the stimulus stops \cite{Mueller:12} and can therefore be ruled
out. In addition, we expect hemodynamic changes to last longer than
a few 100~ms and not to be this highly localized.

Amplitude and timecourse of the signal depended on stimulus intensity,
as its attenuation by OD~0.3 (factor 0.5) and OD~1.3 (factor 0.05)
filters showed (Fig.~\ref{fig:Response}k). The optical path length
of the outer segment rose with a rate independent of the illumination
strength, however, the maximum change increased with stimulus intensity.
The increase rate of most photoproducts such as transducin or phosphodiesterase
is proportional to the actual stimulation strength, as suggested by
kinetic evaluations \cite{Lamb23011996}; consequently, the observed
effect cannot be directly related to these photoproducts. This corroborates
our hypothesis of a secondary more complicated effect being responsible
for this signal. 
\begin{figure}
\centering{}\includegraphics{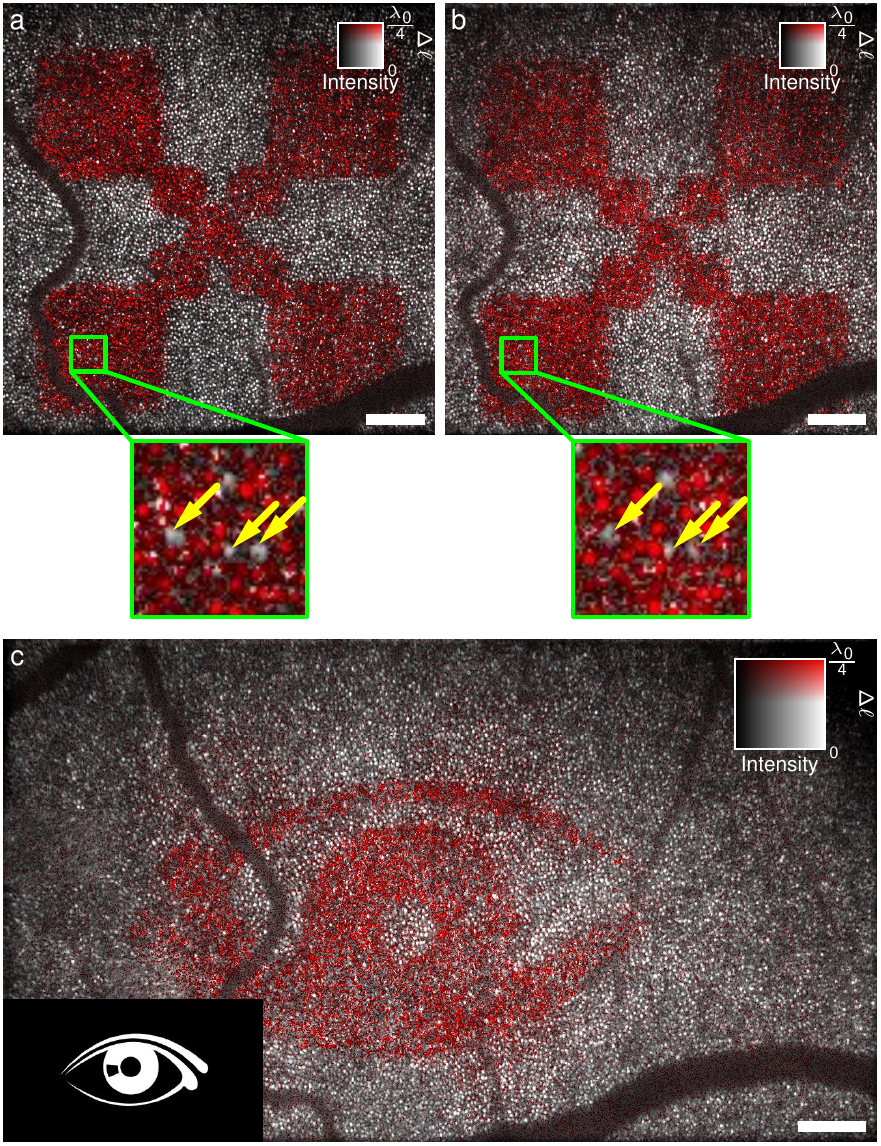}\caption{\textbf{\label{fig:SpatialResponse}After numerical aberration correction
optical path length changes $\Delta\ell$ can be assigned to single
cones.} a,b)~Independent responses to 3~s light stimulus were measured
about 10~min apart. Due to the high lateral resolution individual
photoreceptors can be identified. Here, the same cones did do not
respond to the stimulus (yellow arrows). c)~More complicated stimulation
patterns can be identified, and show which photoreceptors contribute
to an image. The pattern of activated photoreceptors is mirrored compared
to the stimulus pattern. The center wavelength $\lambda_{0}$ of the
OCT was 841~nm. Scale bars are 200~\textmu m.}
\end{figure}

Finally, we enlarged the response by using longer stimuli (Figs.~\ref{fig:Response}g-j,l).
With a permanent stimulus, the increase of the optical path length
of the outer segment was more than 400~nm after 3~s. Here, the optical
path length increased even by more than a quarter of the detection
wavelength, $\lambda_{0}/4\approx210\,\mathrm{nm}$, and the phase
differences exceeded $\pi$ and phase wrapping to $-\pi$ occurred.
This resulted in the visible abrupt change from $+\lambda_{0}/4$
to $-\lambda_{0}/4$ in Figs.~\ref{fig:Response}i,j. Additionally,
in Figs.~\ref{fig:Response}h-j an increased signal background is
visible, which is likely caused by a limited stimulus contrast. 

Using such strong responses, we were able to assign the signals at
the end of the series to single cone photoreceptors; however, a temporal
filter needed to be applied and the phase data had to be unwrapped
to improve signals first. Figure~\ref{fig:SpatialResponse} maps
the response on the en face intensity image. One can see, that although
most cones reacted to the stimulus, some exhibited only a small or
no response. Imaging was sufficiently reproducible to study individual
cones over time, and a measured dataset 10 minutes later showed the
same cones not contributing (Fig.~\ref{fig:SpatialResponse}a,b).
One can also see that in some locations the phase changed abruptly
within a single cone, which might indicate insufficient resolution
to resolve all contributing structures. In the presented field of
view we were only able to see the cones and unresolvable, smaller
rods might have contributed to the observed phases. The same procedure
can also be used for other, more complicated stimulus patterns and
shows which photoreceptors contribute to the image seen by the test
person (Fig.~\ref{fig:SpatialResponse}c).

The key to our observations is certainly the analysis of the optical
path length changes in the outer segments, which was achieved by the
enormous measurement speed and related phase stability of FF-SS-OCT.
We expect further enhancements in the registration and segmentation
of the data to improve the overall signal quality as well. No principle
reason exists why fast scanning OCT systems cannot obtain comparable
results, assuming that sufficient phase stability is established.
The OCT system used in our experiments has a comparably low axial
resolution which is limited by the availability of suitable lasers;
higher axial resolution would result in better separation of retinal
layers, which might not only show a more precise physical origin of
the signal, but would likely also enhance signal quality. Nevertheless,
all results presented here were highly reproducible; measurements
within a few hours gave almost identical timecourses, and only longer
breaks resulted in slightly different overall amplitudes, but showed
qualitatively similar behaviour.

The method presented here resolved small changes within the optical
path length of the outer segment. For backscattered light contributing
to an image, these changes cause a fluctuation in image intensity
as the light interferes. Jonnal et al.~even observed an alternating
reflection signal from individual cones and suggested that this might
be caused by a change in the optical path length within the outer
segment \cite{Jonnal:07}. Depending on resolution and imaging method,
these changes might appear random and in fact previous studies reported
variations in sign and magnitude of observed backscattering intensity
signals \cite{Srinivasan:09,Yao:2015,Teussink:15}. 

Currently, we do not know the cause of the observed signal, and further
research is required to identify the origin or show its pathology-related
value in clinical trials. Nevertheless, detecting the activity of
single cones could be an important tool in early diagnostics of many
diseases that are known to be associated with loss of photoreceptor
function, e.g., age-related macular degeneration \cite{Curcio:1996}
and retinitis pigmentosa \cite{Nagy:2008}. In addition, the functional
loss of single neuronal photoreceptors might also be an indicator
for neurodegenerative diseases, such as Parkinson's \cite{Archibald:2009}
or Alzheimer's disease, e.g., the latter degenerates the retinal ganglion
cells \cite{Williams:2013}, but with the proposed technique one could
also investigate its effect on photoreceptor function.

\section*{Methods}

\paragraph*{Setup}

The setup consisted of a full-field swept-source optical coherence
tomography system (FF-SS-OCT) with an LCD based projection system
to stimulate the retina (Fig.~\ref{fig:Setup}). A cold mirror coupled
the projection light (yellow in Fig.~\ref{fig:Setup}) into the sample
illumination arm of the OCT system (blue in Fig.~\ref{fig:Setup})
and an infrared long pass filter in front of the camera ensured that
no visible light of the stimulation disturbed the OCT system. 

The FF-SS-OCT system used a tunable light source (Superlum Broadsweeper
BS-840-1), which was connected via a fibre coupler to the reference
and to the sample arm of the interferometer. The reference arm illuminated
the camera (Photron FastCAM SA-Z) with a parallel beam through a beam
splitter (green in Fig.~\ref{fig:Setup}), whereas the sample arm
illuminated an area of about $3.6\times1.5\,\mathrm{mm}$ on the retina
through a suitable lens system (blue in Fig.~\ref{fig:Setup}). The
total radiant flux on the retina from the sample illumination was
about $5\,\mathrm{mW}$. The light backscattered by the retina was
imaged onto the camera (red in Fig.~\ref{fig:Setup}) where it was
superimposed with the reference light. An iris controlled the aperture
of the retinal imaging to ensure correct sampling of all lateral frequencies
by the camera.

A single volume was acquired by obtaining multiple images while tuning
the wavelength of the laser. In general, the laser was tuned from
$867$ to $816$ nm, which results in an axial resolution (full width
at half maximum) of 8.5~\textmu m in air, when considering its rectangular
spectrum. During the wavelength scan the camera acquired 380 or 512
images, $640\times368$ pixels each, with 75,000 or 60,000 frames
per second, respectively. Depending on the imaging requirements, acquisition
speed thereby corresponded to 48.8~MHz A-scan-rate or 29.0~MHz A-scan
rate, the maximum volume rates were 180~Hz or 120~Hz, respectively. 

\begin{figure}
\centering{}\includegraphics{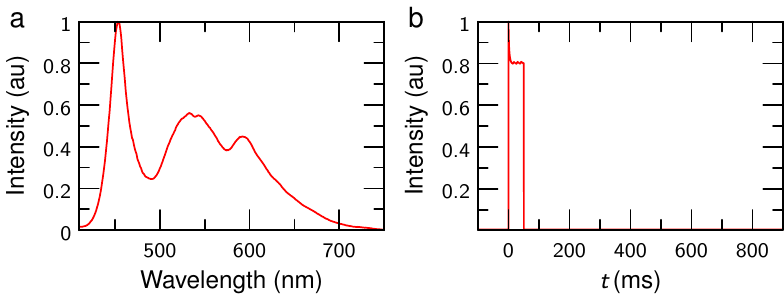}\caption{\textbf{\label{fig:FlashCharacteristics}Characteristics of the flash
used to simulate the retina. }a)~Spectrum of the flash. b)~Intensity
of a 50~ms flash as a function of time, measured using a photodiode. }
\end{figure}
To project the image onto the retina, an LED LCD projector was modified
with an external microprocessor board (Arduino UNO), which controlled
the LED and synchronized the stimulus with the OCT system, i.e., it
triggered the camera and the tunable laser source. The LCD was imaged
via a set of lenses into the conjugate plane of the retina in the
sample illumination arm. Spectrum and timecourse of the white light
retina illumination light were checked carefully and can be found
in Fig.~\ref{fig:FlashCharacteristics}.

Additionally, a passive fixation target and a mesh plastic face mask
were used for optimal imaging. While the fixation target was illuminated
with a green LED to make it visible for the volunteer, care was taken
that no green light of the fixation target was visible in the area
measured by OCT and illuminated by the projector. The mask was custom
fit to the volunteer, ensured a steady and repeatable position of
his head, and made imaged areas highly reproducible.

\paragraph*{Measurement}

The area that can be read out is limited by the lateral resolution
and by the required camera frame rate. Additionally, blood vessels
shadow the outer segment. Therefore, we chose a part of the retina
with few blood vessels, and where cones could be resolved after aberration
correction. Before the first measurement, the volunteer's pupil was
medically dilated and the volunteer was kept in the dark for about
20 minutes, which ensured a reproducible starting point for all measurements,
although measurements with normal light adaption (not presented here)
gave comparable results. Additionally, sufficient time of at least
5~min in between single measurements ensured that results were not
influenced by previous stimulations.

All investigations were done with healthy volunteers; written informed
consent was obtained from all subjects. Compliance with the maximal
permissible exposure (MPE) of the retina and all relevant safety rules
was confirmed by the responsible safety officer. The study was approved
by the ethical board of the University of Lübeck (ethical approval
\textquotedbl{}Ethik-Kommission Lübeck 16-080\textquotedbl{}).

\paragraph*{Pre-processing}

After acquiring the raw images of the camera and before evaluating
the actual phases, the acquired data was pre-processed: First, the
volumetric OCT data for each volume within a time series was reconstructed.
To reduce fixed artifacts within the image, the sum of all datasets
was computed and then subtracted from all volumes within the dataset,
which effectively removed fixed pattern noise as only the retina changes
due to eye motion in between volumes. Afterwards, a Fourier transform
along the time axis for each pixel of the camera reconstructed the
depth information, i.e., intensity and phase of the volume. Lateral
frequency filtering removed noise from outside the aperture and slightly
improved signal-to-noise ratio. 

Second, the images were corrected for dispersion mismatch in reference
and sample arm, as well as intra and inter volume motion. For this,
an optimization algorithm \cite{Wojtkowski:04,Hillmann:2015} removed
the influence of dispersion and axial motion in order to obtain sharp
OCT volumes. All volumes within a series were then shifted in all
three dimensions to maximize their correlation and remove inter volume
displacements. 

In the next step the inner segment/outer segment junction (IS/OS)
of the reconstructed series was segmented independently for each volume.
We shifted all A-scans within each volume axially by using the Fourier
shift theorem such that the IS/OS appeared to be in a single depth.
This ensured reproducible starting conditions for the following step
and the final phase evaluation.

Finally, the volume was corrected for ocular aberrations in the segmented
layer. The aberration correction of the IS/OS layer made single cone
photoreceptors resolvable \cite{Shemonski:15} and was done by an
optimization procedure \cite{Adie:12,Hillmann:2015}. The algorithm
varied Zernike polynomials, which parameterized the aberrations, until
an image quality metric (Shannon's entropy \cite{Fienup:00}) was
optimal. As aberrations varied laterally, the image field was separated
into different areas, which were corrected independently and later
stitched back together. Although this step improved resolution to
resolve individual cones, it was not required to actually observe
the changes in the optical path length within the outer segment.

\paragraph*{Calculation of the optical path length changes from the measured
phases}

The detected phases in FD-OCT do not carry information on absolute
position and to obtain changes, they need to be referenced to phases
in other layers and at other times. To cancel this arbitrary phase
offset in each pixel the reconstructed volumes $f_{xyzt}$ were first
referenced to the last complete volume $f_{xyzt_{0}}$ before the
stimulus began 
\[
f_{xyzt}^{(1)}=f_{xyzt}f_{xyzt_{0}}^{*}\mbox{,}
\]
where $x$, $y$, and $z$ denote the spatial indices of the volume,
and $t$ is the volume number and corresponds to the recording time.
The signal was then averaged over the three layers next to the boundaries
of the outer segment by computing the mean of the complex numbers,
i.e., the mean $\overline{f}^{(1)}$ was computed for the boundaries
of the outer segment, located at $z_{\mathrm{OS,0}}$ and 4 pixels
deeper (corresponding to about 28~\textmu m in air and 20~\textmu m
in tissue) at $z_{\mathrm{OS},1}$ (see also Fig.~\ref{fig:TestStimulation}a).
Finally, the phase difference between both boundaries of the outer
segment was obtained by
\[
f_{xyt}^{(2)}=\overline{f}_{xyz_{\mathrm{OS},0}}^{(1)}\overline{f}_{xyz_{\mathrm{OS},1}}^{(1)*}
\]

The signal quality was improved by applying a lateral Gaussian filter
to the complex data $f_{xyt}^{(2)}$ ($\sigma=16\,\mathrm{px}$, Fig.~\ref{fig:TestStimulation},
Fig.~\ref{fig:Response}) or a moving average filter over 15 volumes
(Fig.~\ref{fig:SpatialResponse}). 

For strong signals, where the phase signal $\Delta\phi_{xyt}=\arg f_{xyt}^{(2)}$
exceeded $\pi$ (Fig.~\ref{fig:SpatialResponse}), the phases were
unwrapped temporally (Fig.~\ref{fig:Response}l and Fig.~\ref{fig:SpatialResponse}).
Phases were finally converted into optical thickness changes using
\[
\Delta\ell=\frac{\Delta\phi}{4\pi}\lambda_{0}\mbox{,}
\]
where $\lambda_{0}$ is the central wavelength of the OCT system,
i.e., $\lambda_{0}\approx841\,\mathrm{nm}$, and $\Delta\phi=\arg f^{(2)}$
is the phase change.

\section*{Acknowledgements}

We would like to thank Reginal Birngruber and Alfred Vogel for valuable
discussions on the results, and Hans-Jürgen Rode, Jörn Wollenzin,
and Christian Winter for their help with the electronics setup. 

This research was sponsored by the German Federal Ministry of Education
and Research (Innovative Imaging \& Intervention in early AMD, contract
numbers 98729873C and 98729873E).

\section*{Author contributions}

D.H.~and G.H.~had the initial idea. D.H. worked on the setup, collected,
analyzed, and interpreted the data, helped obtain funding, and wrote
the manuscript. H.Sp.~worked on the setup and collected data. C.P.~worked
on the setup, collected and interpreted the data. H.Su.~and G.F.~worked
on the setup. G.H.~obtained funding and interpreted the data.

\section*{Competing financial interests}

D.H.~is working for Thorlabs GmbH, which produces and sells OCT systems.
D.H.~and G.H.~are listed as inventors on a related patent application
(application no.~PCT/EP2012/001639). All other authors declare no
competing financial interests.

\bibliographystyle{naturemag}
\bibliography{Optophys}

\begin{thebibliography}{10}
\expandafter\ifx\csname url\endcsname\relax
  \def\url#1{\texttt{#1}}\fi
\expandafter\ifx\csname urlprefix\endcsname\relax\def\urlprefix{URL }\fi
\providecommand{\bibinfo}[2]{#2}
\providecommand{\eprint}[2][]{\url{#2}}

\bibitem{Liebman:74}
\bibinfo{author}{Liebman, P.~A.}, \bibinfo{author}{Jagger, W.~S.} \&
  \bibinfo{author}{M.~W.~Kaplan, F. G.~B.}
\newblock \bibinfo{title}{Membrane structure changes in rod outer segment
  associated with rhodopsin bleaching}.
\newblock \emph{\bibinfo{journal}{Nature}} \textbf{\bibinfo{volume}{251}},
  \bibinfo{pages}{31--36} (\bibinfo{year}{1974}).

\bibitem{Harary:1978}
\bibinfo{author}{Harary, H.}, \bibinfo{author}{Brown, J.} \&
  \bibinfo{author}{Pinto, L.}
\newblock \bibinfo{title}{Rapid light-induced changes in near infrared
  transmission of rods in bufo marinus}.
\newblock \emph{\bibinfo{journal}{Science}} \textbf{\bibinfo{volume}{202}},
  \bibinfo{pages}{1083--1085} (\bibinfo{year}{1978}).

\bibitem{Yao:2015}
\bibinfo{author}{Yao, X.} \& \bibinfo{author}{Wang, B.}
\newblock \bibinfo{title}{Intrinsic optical signal imaging of retinal
  physiology: a review}.
\newblock \emph{\bibinfo{journal}{J. Biomed. Opt.}}
  \textbf{\bibinfo{volume}{20}}, \bibinfo{pages}{090901}
  (\bibinfo{year}{2015}).

\bibitem{Jonnal:07}
\bibinfo{author}{Jonnal, R.~S.} \emph{et~al.}
\newblock \bibinfo{title}{In vivo functional imaging of human cone
  photoreceptors}.
\newblock \emph{\bibinfo{journal}{Opt. Express}} \textbf{\bibinfo{volume}{15}},
  \bibinfo{pages}{16141--16160} (\bibinfo{year}{2007}).

\bibitem{Grieve:2008}
\bibinfo{author}{Grieve, K.} \& \bibinfo{author}{Roorda, A.}
\newblock \bibinfo{title}{Intrinsic signals from human cone photoreceptors}.
\newblock \emph{\bibinfo{journal}{Invest. Ophthalmol. Vis. Sci.}}
  \textbf{\bibinfo{volume}{49}}, \bibinfo{pages}{713} (\bibinfo{year}{2008}).

\bibitem{Srinivasan:09}
\bibinfo{author}{Srinivasan, V.~J.}, \bibinfo{author}{Chen, Y.},
  \bibinfo{author}{Duker, J.~S.} \& \bibinfo{author}{Fujimoto, J.~G.}
\newblock \bibinfo{title}{In vivo functional imaging of intrinsic scattering
  changes in the human retina with high-speed ultrahigh resolution {OCT}}.
\newblock \emph{\bibinfo{journal}{Opt. Express}} \textbf{\bibinfo{volume}{17}},
  \bibinfo{pages}{3861--3877} (\bibinfo{year}{2009}).

\bibitem{Schmoll:2010}
\bibinfo{author}{Schmoll, T.}, \bibinfo{author}{Kolbitsch, C.} \&
  \bibinfo{author}{Leitgeb, R.~A.}
\newblock \bibinfo{title}{In vivo functional retinal optical coherence
  tomography}.
\newblock \emph{\bibinfo{journal}{J. Biomed. Opt.}}
  \textbf{\bibinfo{volume}{15}}, \bibinfo{pages}{041513--041513--8}
  (\bibinfo{year}{2010}).

\bibitem{Teussink:15}
\bibinfo{author}{Teussink, M.~M.} \emph{et~al.}
\newblock \bibinfo{title}{Impact of motion-associated noise on intrinsic
  optical signal imaging in humans with optical coherence tomography}.
\newblock \emph{\bibinfo{journal}{Biomed. Opt. Express}}
  \textbf{\bibinfo{volume}{6}}, \bibinfo{pages}{1632--1647}
  (\bibinfo{year}{2015}).

\bibitem{Huang:1991}
\bibinfo{author}{Huang, D.} \emph{et~al.}
\newblock \bibinfo{title}{Optical coherence tomography}.
\newblock \emph{\bibinfo{journal}{Science}} \textbf{\bibinfo{volume}{254}},
  \bibinfo{pages}{1178--1181} (\bibinfo{year}{1991}).

\bibitem{Bizheva:2006-3693}
\bibinfo{author}{Bizheva, K.} \emph{et~al.}
\newblock \bibinfo{title}{Optophysiology: depth-resolved probing of retinal
  physiology with functional ultrahigh-resolution optical coherence
  tomography}.
\newblock \emph{\bibinfo{journal}{Proc. Natl. Acad. Sci. U. S. A.}}
  \textbf{\bibinfo{volume}{103}}, \bibinfo{pages}{5066--5071}
  (\bibinfo{year}{2006}).

\bibitem{Srinivasan:06}
\bibinfo{author}{Srinivasan, V.~J.}, \bibinfo{author}{Wojtkowski, M.},
  \bibinfo{author}{Fujimoto, J.~G.} \& \bibinfo{author}{Duker, J.~S.}
\newblock \bibinfo{title}{In vivo measurement of retinal physiology with
  high-speed ultrahigh-resolution optical coherence tomography}.
\newblock \emph{\bibinfo{journal}{Opt. Lett.}} \textbf{\bibinfo{volume}{31}},
  \bibinfo{pages}{2308--2310} (\bibinfo{year}{2006}).

\bibitem{Hillmann:11}
\bibinfo{author}{Hillmann, D.}, \bibinfo{author}{L\"{u}hrs, C.},
  \bibinfo{author}{Bonin, T.}, \bibinfo{author}{Koch, P.} \&
  \bibinfo{author}{H\"{u}ttmann, G.}
\newblock \bibinfo{title}{Holoscopy -- holographic optical coherence
  tomography}.
\newblock \emph{\bibinfo{journal}{Opt. Lett.}} \textbf{\bibinfo{volume}{36}},
  \bibinfo{pages}{2390--2392} (\bibinfo{year}{2011}).

\bibitem{Povazay:06}
\bibinfo{author}{Pova\v{z}ay, B.} \emph{et~al.}
\newblock \bibinfo{title}{Full-field time-encoded frequency-domain optical
  coherence tomography}.
\newblock \emph{\bibinfo{journal}{Opt. Express}} \textbf{\bibinfo{volume}{14}},
  \bibinfo{pages}{7661--7669} (\bibinfo{year}{2006}).

\bibitem{Bonin:10}
\bibinfo{author}{Bonin, T.}, \bibinfo{author}{Franke, G.},
  \bibinfo{author}{Hagen-Eggert, M.}, \bibinfo{author}{Koch, P.} \&
  \bibinfo{author}{H\"{u}ttmann, G.}
\newblock \bibinfo{title}{In vivo {Fourier}-domain full-field {OCT} of the
  human retina with 1.5 million {A}-lines/s}.
\newblock \emph{\bibinfo{journal}{Opt. Lett.}} \textbf{\bibinfo{volume}{35}},
  \bibinfo{pages}{3432--3434} (\bibinfo{year}{2010}).

\bibitem{Spahr:15}
\bibinfo{author}{Spahr, H.} \emph{et~al.}
\newblock \bibinfo{title}{Imaging pulse wave propagation in human retinal
  vessels using full-field swept-source optical coherence tomography}.
\newblock \emph{\bibinfo{journal}{Opt. Lett.}} \textbf{\bibinfo{volume}{40}},
  \bibinfo{pages}{4771--4774} (\bibinfo{year}{2015}).

\bibitem{Adie:12}
\bibinfo{author}{Adie, S.~G.}, \bibinfo{author}{Graf, B.~W.},
  \bibinfo{author}{Ahmad, A.}, \bibinfo{author}{Carney, P.~S.} \&
  \bibinfo{author}{Boppart, S.~A.}
\newblock \bibinfo{title}{Computational adaptive optics for broadband optical
  interferometric tomography of biological tissue}.
\newblock \emph{\bibinfo{journal}{Proc. Natl. Acad. Sci. U.S.A.}}
  \textbf{\bibinfo{volume}{109}}, \bibinfo{pages}{7175--7180}
  (\bibinfo{year}{2012}).

\bibitem{Shemonski:15}
\bibinfo{author}{Shemonski, N.~D.} \emph{et~al.}
\newblock \bibinfo{title}{{Computational high-resolution optical imaging of the
  living human retina}}.
\newblock \emph{\bibinfo{journal}{Nature Photon.}}
  \textbf{\bibinfo{volume}{9}}, \bibinfo{pages}{440--443}
  (\bibinfo{year}{2015}).

\bibitem{Hillmann:2015}
\bibinfo{author}{Hillmann, D.} \emph{et~al.}
\newblock \bibinfo{title}{Aberration-free volumetric high-speed imaging of in
  vivo retina. {Submitted to \emph{Sci.~Rep.}}}  (\bibinfo{year}{2015}).

\bibitem{Uhl1977}
\bibinfo{author}{Uhl, R.}, \bibinfo{author}{Hofmann, K.~P.} \&
  \bibinfo{author}{Kreutz, W.}
\newblock \bibinfo{title}{Measurement of fast light-induced disc shrinkage
  within bovine rod outer segments by means of a light-scattering transient}.
\newblock \emph{\bibinfo{journal}{Biochim. Biophys. Acta}}
  \textbf{\bibinfo{volume}{469}}, \bibinfo{pages}{113--122}
  (\bibinfo{year}{1977}).

\bibitem{Mueller:12}
\bibinfo{author}{Müller, H.~H.} \emph{et~al.}
\newblock \bibinfo{title}{{Imaging thermal expansion and retinal tissue changes
  during photocoagulation by high speed OCT}}.
\newblock \emph{\bibinfo{journal}{Biomed. Opt. Express}}
  \textbf{\bibinfo{volume}{3}}, \bibinfo{pages}{1025--1046}
  (\bibinfo{year}{2012}).

\bibitem{Lamb23011996}
\bibinfo{author}{Lamb, T.~D.}
\newblock \bibinfo{title}{Gain and kinetics of activation in the {G}-protein
  cascade of phototransduction}.
\newblock \emph{\bibinfo{journal}{Proc. Natl. Acad. Sci. U.S.A.}}
  \textbf{\bibinfo{volume}{93}}, \bibinfo{pages}{566--570}
  (\bibinfo{year}{1996}).

\bibitem{Curcio:1996}
\bibinfo{author}{Curcio, C.~A.}, \bibinfo{author}{Medeiros, N.~E.} \&
  \bibinfo{author}{Millican, C.~L.}
\newblock \bibinfo{title}{Photoreceptor loss in age-related macular
  degeneration.}
\newblock \emph{\bibinfo{journal}{Invest. Ophthalmol. Vis. Sci.}}
  \textbf{\bibinfo{volume}{37}}, \bibinfo{pages}{1236} (\bibinfo{year}{1996}).

\bibitem{Nagy:2008}
\bibinfo{author}{Nagy, D.}, \bibinfo{author}{Sch\"onfisch, B.},
  \bibinfo{author}{Zrenner, E.} \& \bibinfo{author}{J\"agle, H.}
\newblock \bibinfo{title}{Long-term follow-up of retinitis pigmentosa patients
  with multifocal electroretinography}.
\newblock \emph{\bibinfo{journal}{Invest. Ophthalmol. Vis. Sci.}}
  \textbf{\bibinfo{volume}{49}}, \bibinfo{pages}{4664} (\bibinfo{year}{2008}).

\bibitem{Archibald:2009}
\bibinfo{author}{Archibald, N.~K.}, \bibinfo{author}{Clarke, M.~P.},
  \bibinfo{author}{Mosimann, U.~P.} \& \bibinfo{author}{Burn, D.~J.}
\newblock \bibinfo{title}{The retina in {Parkinson{\textquoteright}s} disease}.
\newblock \emph{\bibinfo{journal}{Brain}} \textbf{\bibinfo{volume}{132}},
  \bibinfo{pages}{1128--1145} (\bibinfo{year}{2009}).

\bibitem{Williams:2013}
\bibinfo{author}{Williams, P.~A.} \emph{et~al.}
\newblock \bibinfo{title}{Retinal ganglion cell dendritic degeneration in a
  mouse model of {Alzheimer's} disease}.
\newblock \emph{\bibinfo{journal}{Neurobiol. Aging}}
  \textbf{\bibinfo{volume}{34}}, \bibinfo{pages}{1799 -- 1806}
  (\bibinfo{year}{2013}).

\bibitem{Wojtkowski:04}
\bibinfo{author}{Wojtkowski, M.} \emph{et~al.}
\newblock \bibinfo{title}{Ultrahigh-resolution, high-speed, {Fourier} domain
  optical coherence tomography and methods for dispersion compensation}.
\newblock \emph{\bibinfo{journal}{Opt. Express}} \textbf{\bibinfo{volume}{12}},
  \bibinfo{pages}{2404--2422} (\bibinfo{year}{2004}).

\bibitem{Fienup:00}
\bibinfo{author}{Fienup, J.~R.}
\newblock \bibinfo{title}{Synthetic-aperture radar autofocus by maximizing
  sharpness}.
\newblock \emph{\bibinfo{journal}{Opt. Lett.}} \textbf{\bibinfo{volume}{25}},
  \bibinfo{pages}{221--223} (\bibinfo{year}{2000}).

\end{thebibliography}

\end{document}